\documentclass[a4paper,11pt]{article}
\pdfoutput=1 
\usepackage{jinstpub}

\usepackage{enumitem, xspace, multirow, adjustbox, rotating}

\newcommand\superbit{\textsc{SuperBIT}\xspace}
\newcommand\drs{\textsc{drs}\xspace}
\newcommand\hab{\textsc{hab}\xspace}
\newcommand\csa{\textsc{csa}\xspace}
\newcommand\cnes{\textsc{cnes}\xspace}
\newcommand\nasa{\textsc{nasa}\xspace}
\newcommand\gnss{\textsc{gnss}\xspace}

\title{Download by Parachute:\\ Retrieval of Assets from High Altitude Balloons}

\author[a,1]{E.~L.~Sirks,\note{Corresponding author.}}
\author[b]{P.~Clark,}
\author[a,b,c]{R.~J.~Massey,}
\author[d]{S.~J.~Benton,}
\author[b]{A.~M.~Brown,}
\author[e]{C.~J.~Damaren,}
\author[f]{T.~Eifler,}
\author[d]{A.~A.~Fraisse,}
\author[a]{C.~Frenk,}
\author[g]{M.~Funk,}
\author[h]{M.~N.~Galloway,}
\author[i,j]{A.~Gill,}
\author[k]{J.~W.~Hartley,}
\author[e, j]{B.~Holder,}
\author[l]{E.~M.~Huff,}
\author[a,c]{M.~Jauzac,}
\author[d]{W.~C.~Jones,}
\author[a, c]{D.~Lagattuta,}
\author[i,j]{J.~S.-Y.~Leung,}
\author[d]{L.~Li,}
\author[d]{T.~V.~T.~Luu,}
\author[l]{J.~McCleary,}
\author[j]{J.~M.~Nagy,}
\author[i,j,k]{C.~B.~Netterfield,}
\author[d]{S.~Redmond,}
\author[l]{J.~D.~Rhodes,}
\author[d]{L.~J.~Romualdez,}
\author[b]{J.~Schmoll,}
\author[j,k]{M.~M.~Shaaban,}
\author[a]{and S.-I.~Tam}

\affiliation[a]{Institute for Computational Cosmology, Durham University,\\South Road, Durham DH1 3LE, UK}
\affiliation[b]{Centre for Advanced Instrumentation, Durham University,\\South Road, Durham DH1 3LE, UK}
\affiliation[c]{Centre for Extragalactic Astronomy, Department of Physics, Durham University,\\Durham DH1 3LE, UK}
\affiliation[d]{Department of Physics, Princeton University,\\Jadwin Hall, Princeton, NJ, USA}
\affiliation[e]{University of Toronto Institute for Aerospace Studies (UTIAS),\\4925 Dufferin Street, Toronto, ON, Canada}
\affiliation[f]{Department of Astronomy/Steward Observatory,\\933 North Cherry Avenue, Tucson, AZ 85721-0065, USA}
\affiliation[g]{Space Markt,\\124 Route de Chêne, Geneva, CH 1224, Switzerland}
\affiliation[h]{Institute of Theoretical Astrophysics, University of Oslo,\\Blindern, Oslo, Norway}
\affiliation[i]{Department of Astronomy and Astrophysics, University of Toronto,\\50 St. George Street, Toronto, ON, Canada}
\affiliation[j]{Dunlap Institute for Astronomy and Astrophysics, University of Toronto,\\50 St. George Street, Toronto, ON, Canada}
\affiliation[k]{Department of Physics, University of Toronto,\\60 St. George Street, Toronto, ON, Canada}
\affiliation[l]{Jet Propulsion Laboratory, California Institute of Technology,\\4800 Oak Grove Drive, Pasadena, CA, USA}

\emailAdd{ellen.l.sirks@durham.ac.uk}

\abstract{We present a publicly-available toolkit of flight-proven hardware and software to retrieve 5\,TB of data or small physical samples from a stratospheric balloon platform. 
Before launch, a capsule is attached to the balloon, and rises with it. 
Upon remote command, the capsule is released and descends via parachute, continuously transmitting its location.
Software to predict the trajectory can be used to select a safe but accessible landing site.
We dropped two such capsules from the \superbit telescope, in September 2019.
The capsules took $\sim$37\,minutes to descend from $\sim$30\,km altitude.
They drifted 32\,km and 19\,km horizontally, but landed within 300\,m and 600\,m of their predicted landing sites. 
We found them easily, and successfully recovered the data. 
We welcome interest from other balloon teams for whom the technology would be useful.}

\keywords{Balloon instrumentation, Data handling, Data compression, Models and simulations, Large detector-systems performance}

\begin{document}
\maketitle
\flushbottom

\section{Introduction}
High altitude balloon (\hab) missions are increasing in number, duration, and expense. 
Some acquire enough data that transmitting it to the ground would be impossible due to limited band-width or cost; others acquire physical samples that must be returned to the ground for full analysis.
Mid-flight retrieval could improve a mission's efficiency, by using early results to optimise later data acquisition.
Retrieval at any time mitigates the critical risk of total loss if the main hardware were damaged upon landing or lost, e.g.\ at sea.

Examples of small balloons include the $\sim2000$ radiosondes launched every day for weather forecasting, as well as instruments flown by amateur groups for scientific or educational purposes.
Less than 20\% of the $\sim$US\textdollar200 radiosondes launched in the USA are recovered, which prohibits upgrades to $\sim$US\textdollar1000 ozonesondes \citep{a}, or increases in the number of weather stations, whose sparsity in the Southern hemisphere particularly limits forecasting precision \citep{b}.

An example of a large scale \hab mission is the Superpressure Balloon-borne Imaging Telescope \citep[\superbit;][]{c,d}. \superbit is an astronomical telescope that rises above 99\% of the Earth's turbulent atmosphere to achieve stabilised \citep{e, f} high-resolution imaging at visible and near-UV wavelengths, with a field of view 36 times larger than the Hubble Space Telescope's Advanced Camera for Surveys/Wide Field Camera. \superbit is currently scheduled for a 50--100 day long duration flight, during which it will obtain $\sim$50\,GB of uncompressed science data per day; a successor is already being designed that will obtain 20 times more \citep{g}. 

Line-of-sight radio communications can achieve 100\,Mbps but, on a long duration flight, global satellite communication systems are limited to 1\,Mbps (10.5\,GB per day), which is not exclusively used for image transfer, and cost up to US\$0.50 per MB.\footnote{See \url{https://www.mailasail.com/Communication/Iridium-Pilot-Airtime}}

We have developed the \superbit Data Recovery System (\drs) to recover assets from any balloon, any time it is over land. In default configuration, each \drs capsule includes 5\,TB of storage, accessible over wifi ethernet. These are attached to a \hab platform before launch, and ascend as usual. Following a remote command, they descend via parachute, transmitting their location via Iridium message -- and continuing to transmit as well as beep audibly after landing. We have also calibrated and tested software to predict the descent trajectory and landing site. This software helps to optimise the moment of release, so the \drs lands safely but accessibly, and assists retrieval on the ground. We successfully used two \drs capsules during {\superbit}'s science commissioning test flight, and intend to use several more during its long duration mission. We also welcome interest from other \hab mission teams for whom the technology would be useful.

This paper is organized as follows. Section~\ref{sec:reqs} details safety and other requirements. Section~\ref{sec:hardware} describes the \drs hardware and its release mechanism. Section~\ref{sec:software} describes the algorithm we use to predict its landing site. Section~\ref{sec:test} describes an end-to-end test of the \drs during the 2019 \superbit commissioning flight. We draw conclusions, and outline plans for future improvements in section~\ref{sec:conclusion}.

\newpage
\section{Requirements}\label{sec:reqs}
This section summarises the main safety requirements for a \drs to be allowed to be jettisonned from a balloon launched by the Canadian Space Agency (\csa) and Centre National d'\'Etudes Spatiales (\cnes) from the Timmins Stratospheric Balloon Base in Ontario, Canada in September 2019. The requirements were set in conjunction with the International Civil Aviation Organization's Convention on International Civil Aviation Rules of the Air (Annex 2), but note that requirements may differ at other launch sites or for other agencies.

Relevant safety requirements include (but are not limited to)

\begin{enumerate}[label=(R\arabic*), leftmargin=*]
\item \emph{Electrical safety}: \label{req:electrical}
To prevent risk of fire, the gondola and/or \drs must be equipped with a fuse. All cables must be rated for a current greater than the fuse, and must also be insulated, protected, and secured. Electrical connectors must be designed so that there is no ambiguity in their connection. Static charges must be drained away.
\item \emph{Mechanical safety}: \label{req:mechanical}
The \drs capsule must not detach from the \hab platform unless commanded. In particular, the release mechanism must be sufficiently robust to withstand shocks during launch and descent (in case it is not released). The maximum vertical and horizontal acceleration for a 750\,kg payload on a 14 million cubic foot zero-pressure balloon are $6.4g$ (vertical) and $1.3g$ (horizontal), which occur during parachute deployment.\footnote{According to \cnes internal document BSO-MU-0-4793-CN-VA.} We add these in quadrature, with a safety factor of $\times2$, and adopt a requirement on the \drs to withstand accelerations up to $13g$. 
\item \emph{Control of Fault Propagation}: \label{req:twostep}
Two or more active steps must be taken by an operator to initiate the release of a \drs capsule. In the event of power failure, there must be no change in the state of any safety barrier, and systems must switch to safe mode. It must not be possible for an electrical circuit to be activated as a result of an action on any other circuit, or through the effect of external events.
\item \emph{Descent safety}: \label{req:vertical_speed}
As the \drs reaches ground level, it must have vertical speed 
\begin{equation}
|v_{z}| < \left( 5 + \frac{3.4\,\mathrm{kg}}{m} \right)\mathrm{m\,s}^{-1}\,,
\end{equation}
where $m$ is its mass.\footnote{Equation provided by \csa.} This safety criterion applies to any package with total mass $<2\,$kg and areal density $<13\,$g\,cm$^{-2}$, defined as the mass of the package divided by the area of its smallest surface.
\end{enumerate}

\noindent To be useful, the \drs must also meet several practical requirements

\begin{enumerate}[label=(R\arabic*), leftmargin=*, resume]
\item \emph{Easy to find}: \label{req:findable} The \drs must be easy to find after landing, visibly and audibly. 
\item \emph{Labelling}: \label{req:label} In case the \drs is found by a person not associated with the \hab mission, it must be labelled with a safety warning about the electrical hazards, and contact details for more information or where to return the capsule.
\item \emph{Predictable}: \label{req:software} It must be possible to predict the descent trajectory and landing site of the \drs within 5\,km (requirement) or 1\,km (goal), in order to make go/no-go decisions about release. More accurate performance will open more potential landing sites that avoid e.g.\ towns and lakes, and cluster near remote roads to aid recovery. This code must run in $<30\,$s, so that accurate decisions can be made about the timing of release from even a fast-moving \hab. A slower but more accurate prediction may also be useful to assist recovery, in the event of communication loss.
\end{enumerate}

\section{Hardware}\label{sec:hardware}

\begin{figure}[t]
\centering
\includegraphics[width=\textwidth]{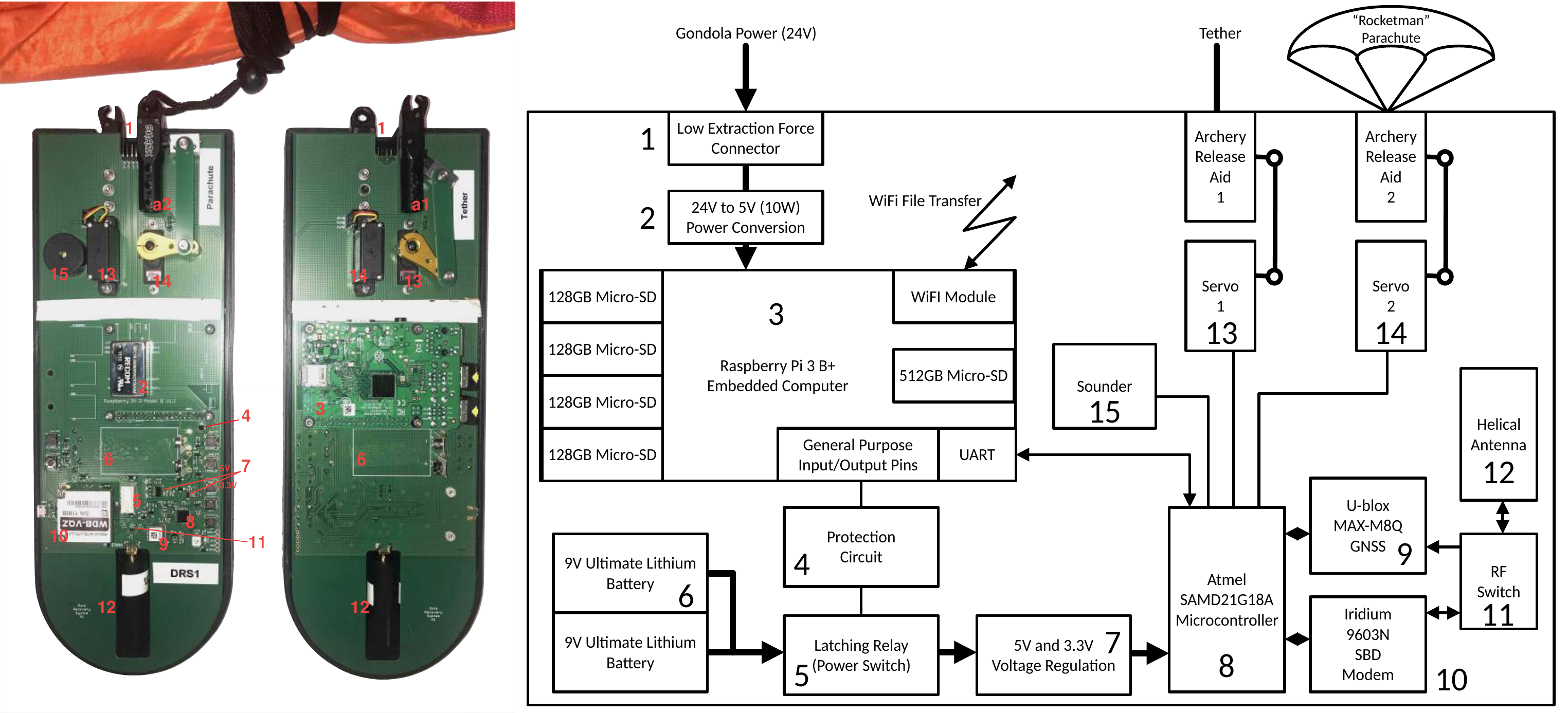}
\caption{Left: Front side of the PCB. Middle: Rear side of the PCB. The red numbers refer to the numbers on the block diagram. Right: Block diagram of the PCB. The \textbf{a1} and \textbf{a2} indicate the archery release mechanisms 1 and 2 respectively.}
\label{fig:PCB}
\end{figure}

The \drs hardware design and operations software are open source.\footnote{Available, with full operating instructions, from \url{https://github.com/PaulZC/Data\_Recovery\_System}.} All components are integrated onto on a custom 300\,mm$\times$100\,mm printed circuit board (PCB). Throughout this section, numbers in curly brackets refer to component labels in figure~\ref{fig:PCB}.

The main function of the \drs is to carry large quantities of science data to the ground and allow its recovery. It is, in effect, 'remote storage with benefits' for the main data acquisition computer (IFC). Data could be transferred into that remote storage either over a wired interface, such as USB or Ethernet, or wirelessly. In the case of \superbit, the IFC and the \drs are physically separated, making USB an unwise choice as, e.g., USB2.0 has a maximum cable length of 5\,m. We selected wireless rather than wired Ethernet in order to avoid having to use an 8-way connector, although our experience with low extraction force connectors since then has suggested that an Ethernet interface would work well.

The IFC manages many tasks, such as command forwarding, telemetry downlink, and science camera housekeeping. It is essential that the file transfer into the \drs does not take resources from those operations since the IFC is the gateway to the rest of the \superbit payload. Using a Raspberry Pi single board computer in the \drs allowed us to implement a wireless or Ethernet interface in a straight forward way. It also simplified the mirroring of files from the IFC into \drs storage, by having essentially a Unix computer at both ends of the transfer. 

\subsection{Enclosure}

The PCB is protected by a 3D-printed ABS-like cover, which is manufactured in two identical halves and sealed around the lower two thirds to limit water ingress (with a moisture barrier vent to allow pressure equalisation).\footnote{The material is similar to ABS, but is a bit easier to work with and does not suffer from the same delamination problems. See \url{https://e3d-online.com/spoolworks-edge}.} This is enclosed inside a softer outer shell, made from moulded expanding polyurethane (PU) foam. Nylon paracord of diameter 2.4\,mm is embedded into the foam, so it can be tied over the top of the cover to secure it; and a nylon sheet lining the mould forms a smooth outer surface on which warnings and contact details can be written in permanent marker \ref{req:label}. The entire \drs, including parachute and batteries, weighs 1029\,grams and has areal density $5.8\,$g\,cm$^{-2}$.

\subsection{Power}

The \drs capsule will be powered down during most of the \hab mission. This prevents accidental or erroneous release. 
When the \drs is required, remotely switched (and fused) 12--48\,V DC power is supplied from the gondola, via a low extraction force connector \{1\} with three pins arranged symmetrically and with redundancy on ground \ref{req:electrical}. A medical-grade, switch-mode DC-DC converter \{2\} regulates power to 5\,V. The embedded Raspberry Pi computer \{3\} automatically boots up, enables its Wi-Fi$^{\rm TM}$ network, and connects to the main gondola flight computer. In its current configuration, the \drs uses a power cable with only 3 pins, to minimise the force required to disconnect. Further tests have shown that a connector with 8 pins (arranged in an asymmetric configuration to meet requirement \ref{req:electrical}) will also work, so future versions of the \drs may use wired Ethernet with Power-over-Ethernet.

Immediately before descent, a latching power relay \{5\} is switched, and two Energizer Ultimate Lithium 9V (PP3) batteries \{6\} supply similarly regulated \{7\} power to a tracking subsystem \{8--15\}. These batteries will henceforth remain powered, and are the only components of the jettisoned \drs that could be considered potentially hazardous \ref{req:electrical}. However, they are compliant with safety test criteria T1--T8 defined in Section 38.3 of Ref. \cite{h}, which include transportation safety and altitude simulation. Indeed, we have used these batteries without incident in $>30$ \hab flights \citep{i}. 

\subsection{Raspberry Pi}

The Raspberry Pi provides the front-end user interface for the \drs, accessible during the mission via {\tt ssh} from the main gondola flight computer. For \superbit, it is also the heart of the `recoverable assets', hosting up to 5\,TB of solid-state data storage (1\,TB micro SD card that includes the operating system, plus $4\times1$\,TB micro SD cards, through 480\,MB\,s$^{-1}$ USB2.0). Data can be copied to this at any time before release, using gondola power. As a useful backup in case of faults e.g.\ due to cosmic rays in the space-like environment, data is constantly uploaded instead of all at once right before release.

We have also considered using the Raspberry Pis to pre-process and analyse science data during flight, but found they overheated when used for long durations in vacuum and inside the PU foam enclosure: implementing this would require thermal redesign. 

\begin{figure}[t]
\centering
\includegraphics[width=0.8\textwidth]{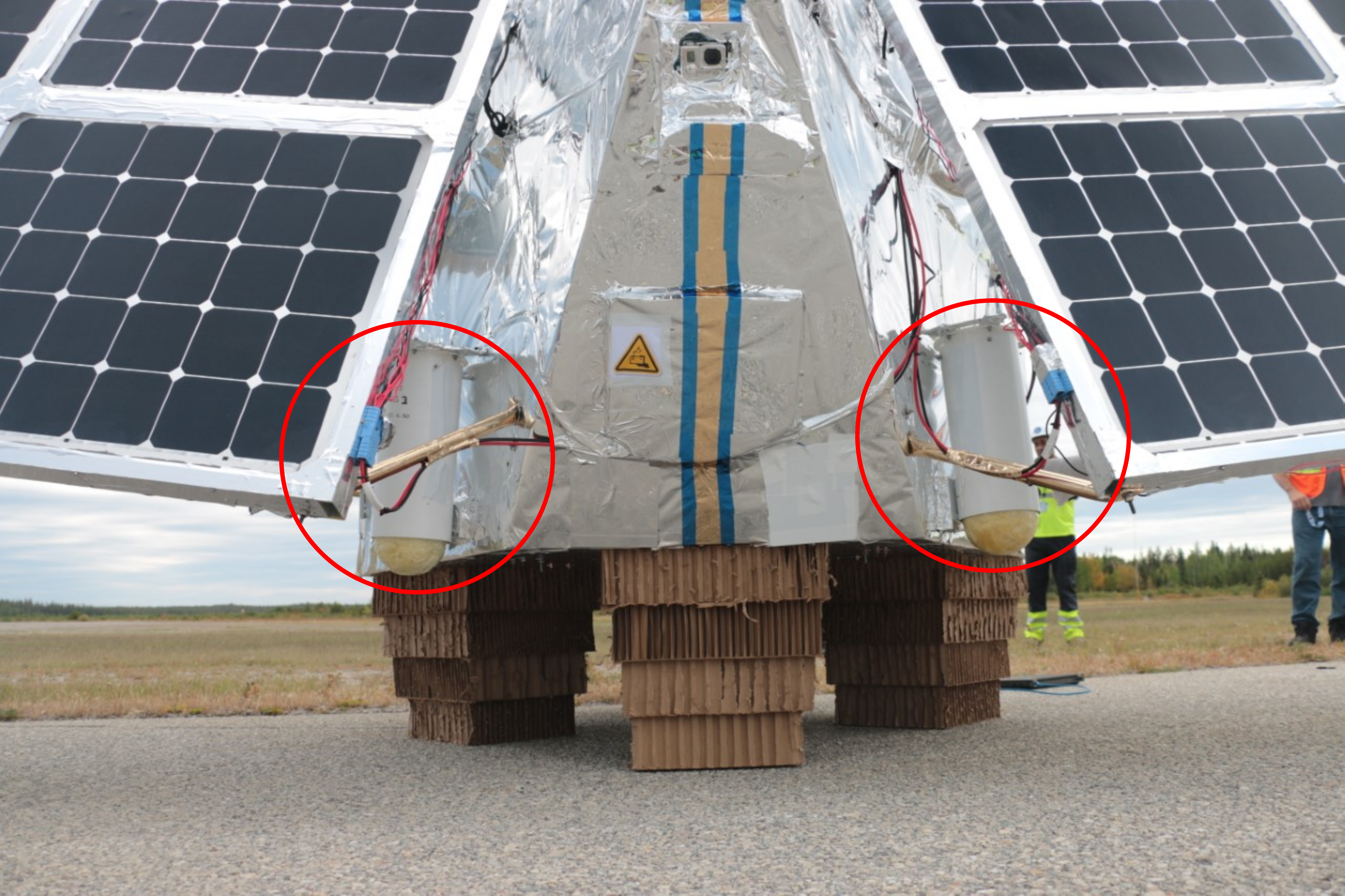}
\caption{Two \drs capsules (highlighted by red circles), mounted on the back of the \superbit telescope just before launch on September 17, 2019. The white launch tubes stay attached to the telescope when the capsules are dropped. The PU foam surrounding the circuit boards can be seen protruding from the bottom of the tubes. 
The cardboard crush pads underneath \superbit are intended to soften impact upon landing.}
\label{fig:superbit+capsules}
\end{figure}

\subsection{Release Mechanism}

Each \drs capsule is packaged inside a plastic drainpipe (diameter 150\,mm, length 350\,mm), to limit swinging and to constrain the parachute before release (figure \ref{fig:superbit+capsules}). These `launch tubes' remain attached to the gondola after the \drs is released.

Inside each tube is a short power cable and a loop of 2.4\,mm diameter nylon paracord. As with our balloon tracking payload \citep{i}, the \drs grips the loop using a sprung release-aid mechanism developed for archery, and operated here via a servo \{13\} stripped, cleaned and re-lubricated with `space-grease' (Castrol's Braycote 601 EF). The strength of the release mechanism was tested against requirement~\ref{req:mechanical} by holding the PCB upside-down and hanging 13\,kg of lead bricks from the nylon cord. The release mechanism held, and no damage to the PCB or nylon cord was observed.

\subsection{Two-Step Instructions for Release}

Two further actions are required to release the \drs\ \ref{req:twostep}, once the Raspberry Pi is powered up.
First, the ground team must {\tt ssh} into the Raspberry Pi and run a `Power On' python script, which configures its GPIO pins to switch on the latching relay \{5\}. A discrete logic protection circuit \{4\} requires three of the GPIO pins to be in the correct state before the relay is triggered. The pins and states have been selected to prevent the relay from being accidentally triggered as the Pi goes through its boot process. Once the relay is triggered, the \drs's internal batteries power the microcontroller \{8\}, which goes through its own start-up procedure and starts to monitor its serial (UART) port for a `Go' command. The Global Navigation Satellite System (\gnss) receiver \{9\} is also powered up and starts to establish a fix. The \gnss NMEA messages are sent through the serial port of the microcontroller and logged by the Raspberry Pi. This can be monitored and, if required, the drop can be delayed until it is confirmed that the \gnss has established a fix. 

Second, the ground team must use {\tt ssh} to run another python script that sends a `Go' command to the microcontroller via its serial (UART) port, then immediately shuts down the Raspberry Pi. 30 seconds later (time for the Pi to shut down gracefully), the microcontroller enables 5\,V power to the servo via a P-channel FET then generates the correct Pulse Width Modulated (PWM) signal to move the servo to the open position. As the \drs is released, the low extraction force connector pulls apart, disconnecting power to the Raspberry Pi, which will remain inactive until recovery. 
If the `Go' script is accidentally run before the first `Power On' script, the script will have no effect as the microcontroller will be unpowered and the `Go' command ignored. 
If either of the microcontroller actions fail, e.g.\ due to its code crashing, the release will not open. 

\subsection{Parachute}\label{sec:parachute}

The parachute is initially folded on top of the \drs, inside the plastic launch tube (figure~\ref{fig:superbit+capsules}). It unfolds when the capsule slides out of the white tube. 
We use a 4\,foot (1.22\,m) Rocketman parachute, which is expected to slow the descent of our 1029\,g payload to terminal velocity $<4\,$m\,s$^{-1}$ at ground level, easily meeting requirement~\ref{req:vertical_speed}.\footnote{See \url{https://the-rocketman.com/recovery-html/}.} It is coloured bright orange, to aid recovery on the ground \ref{req:findable}.\footnote{Optionally, a second servo \{14\} and archery release can be used to release the parachute once it has been confirmed to have reached the ground. This option could prevent the \drs from being dragged by the parachute, or allow it to fall to the ground if the parachute has become caught in a tree. However, it introduces a risk of the parachute being released prematurely, through human error. To militate against this risk, the second release can only be opened by sending a Mobile Terminated (MT) SBD message containing a time code. The microcontroller will only respond if the time code matches \gnss time to within an appropriate interval; it will ignore (and delete) all other messages, so old queued or erroneous MT messages have no effect.}

\subsection{Tracking and recovery}\label{sec:tracking}

During descent and after landing, communication is maintained with the \drs via Iridium 9603N satellite modem \{10\}. The microcontroller alternately switches \{11\} between monitoring its location vis \gnss then transmitting this information via Mobile Originated Iridium SBD messages. A large, helical antenna \{12\} is shared for these tasks, saving weight while achieving superior performance than a patch antenna, especially after landing horizontally on ground, in trees or on water. A small Radio Frequency (RF) switch is used to connect the antenna to either the \gnss or the Iridium modem. The switch shields the \gnss during Iridium transmit bursts. This subsystem is a modified version of Ref. \cite{i}'s \hab tracking toolkit.

A sounder \{15\} begins beeping after the `Go' command is received. Thus a recovery crew can head to \gnss coordinates (in a worst case, transmitted immediately before landing), then look for a bright orange parachute and listen for beeps \ref{req:findable}. The sounder can be disabled (or re-enabled), and the frequency with which the \drs reports its location can be adjusted, via Iridium MT message to the \drs. Depending on this frequency, the batteries have an expected operating lifetime of 2--6\,weeks. Electrical hazard warnings and contact information written on the nylon surface in permanent marker are easily visible after this time, even in wet conditions \ref{req:label}.

\section{Software to Predict Descent Trajectories}\label{sec:software}

The key remaining requirement \ref{req:software} is software to quickly and accurately predict the landing site of the \drs.
We have adapted open source python code, originally written to simulate the trajectories of tropospheric sounding balloons.\footnote{\url{https://github.com/pnuu/pyBalloon} by Panu Lahtinen, currently at the Finnish Meteorological Institute.} Such trajectories included an ascent phase on a weather balloon and a descent phase of the payload on a parachute. We are principally interested in the descent phase, and have improved and calibrated its accuracy. The code remains open source.\footnote{See \url{https://github.com/EllenSirks/pyBalloon}.}

\subsection{Data}\label{sec:data}
\subsubsection{Weather Models}\label{sec:weather_data}

We use Global Forecast System (GFS) weather models produced by the National Centers for Environmental Prediction (NCEP). They are generated every six hours, at 00:00, 6:00, 12:00, and 18:00 GMT, then become publicly available $\sim$$3.5$\,hours later (for current weather conditions) to 5\,hours later (for a forecast up to 16\,days into the future).\footnote{See \url{https://www.ncdc.noaa.gov/data-access/model-data}.}

The forecasts include air density, temperature, wind speeds, and geopotential heights in voxels across the globe, with a horizontal resolution of 0.5 degrees, and at 34 air pressure levels, ranging from 1000\,mb (low altitude) to 0.4\,mb (high altitude).\footnote{GFS models are calculated at air pressure levels: 0.4, 1, 2, 3, 5, 7, 10, 15, 20, 30, 40, 50, 70, 100, 150, 200, 250, 300, 350, 400, 450, 500, 550, 600, 650, 700, 750, 800, 850, 900, 925, 950, 975, and 1000\,mb.} The geopotential heights represent the height above sea level of a given pressure level.\footnote{See `height' at  \url{https://w1.weather.gov/glossary}.} This is an estimated height based on temperature and pressure data. At relatively low altitudes, the geopotential height is approximately equal to the geometric height. E.g. at the \superbit flight altitude, $\sim$30\,km, the difference is less than 150\,m. The models have a vertical resolution between $\sim$$200$\,m near ground level to $5$\,km at stratospheric altitudes ($\sim$$50$\,km).

Conditions are forecast with a time resolution of 3\,hours. The difference between the production time of forecasts and the trajectory time has a large effect on our accuracy, and so we introduce variable $t_\mathrm{future}$, the number of hours a forecast is predicting into the future. For example, for conditions at 16:00, the forecast nearest in time is produced at 12:00 with $t_{\mathrm{future}}=3$\,hours.
An ensemble of weather forecasts, generated from slightly perturbed initial conditions, are also available for 9\,days (after which their files are deleted, and the main model is moved to archival storage). We have experimented using the ensemble forecasts to estimate uncertainty -- but find their variance to be smaller than other sources of uncertainty in our calculations, and cannot access them for historic flights, so do not exploit them.

We require a look-up table of atmospheric conditions at higher resolution than the GFS forecasts. We shall therefore interpolate all variables in vertical columns using a cubic B-spline, in latitude and longitude using bilinear interpolation, then linearly in time. Compared to this scheme, nearest neighbour interpolation degrades the accuracy of our landing site predictions by 
28\% (4\% from spatial interpolation and 23\% from temporal interpolation).

\subsubsection{Altitude of the ground}\label{sec:altitude_data}
For locations with a latitude between -60 and 60 degrees, we use tables of ground altitude as a function of latitude and longitude with a resolution of 1 arcsecond or approximately 30\,m at the equator.\footnote{See \url{http://srtm.csi.cgiar.org/srtmdata/}.} For any other latitudes we use tables with slightly lower resolution of 3 arcseconds as the high resolution data are not available for these regions.\footnote{See \url{http://viewfinderpanoramas.org/}.} At a given location, we assign the altitude of the closest grid point in the tables as the elevation.

\subsubsection{Test Flights}\label{sec:test_flights}

We have access to the trajectories of 30 flights in which a real payload ascended by weather balloon then descended via parachute \citep{i}. These took place between 2018 and 2019, in Switzerland (20), Greenland (4), and Morocco (6), and are listed in table~\ref{tab:funkflights}. During each flight, the longitude, latitude, and altitude of the payload was recorded by \gnss in $\sim$$5$\,minute intervals.

We exploit these trajectories to calibrate our software and test its accuracy; however, they were not originally intended for this purpose. For example, the payload mass was $\sim$$1.6$\,kg (and always $<2$\,kg for legal reasons) but not accurately recorded on each occasion. The parachute was  a 7\,foot (2.13\,m) Rocketman parachute, of the same design as our DRS but larger. Furthermore, the time at which the balloon burst was not recorded (even though it was detected via the on-board accelerometer). At the highest point recorded by \gnss, the payload could be either ascending or descending. It was only guaranteed to be descending at the time and location recorded \emph{after} the highest point. We therefore use this as the initial condition for descent trajectories.

\begin{sidewaystable}
\centering
\footnotesize
\caption{\footnotesize Descent trajectories of real payloads, logged via \gnss. We use the top 30 descents to calibrate and test our software. During each of these flights, a payload ascended via weather balloon, then descended via parachute. The exact cutdown point was not recorded, so we list the time, date and location of the first \gnss location after its highest: the first moment at which the payload is guaranteed to be descending. The landing site is the mean position of the \gnss locations recorded with the same altitude. The range $d$ is the straight-line horizontal distance between the release point and landing site. $\langle t_{\rm future} \rangle$ is the mean $t_{\rm future}$ of all the forecasts used at each altitude step in a given predicted trajectory. The bottom two rows show the position and time of release of the \drs capsules launched with \superbit for an end-to-end demonstration. The flights marked by an asterisk had limited data logs or extreme trajectories, and were excluded from our statistical analysis.}
\smallskip
\tabcolsep=0.11 cm
\begin{tabular}{|cccccccccccccc|}
\hline
\multirow{2}{*}{Date} & \multirow{2}{*}{Time} & \multirow{2}{*}{Location} & \multicolumn{3}{c}{Release point} & \multicolumn{2}{c}{Landing site} & Time of & Horizontal & $\langle t_{\rm future}\rangle$ &  \multicolumn{2}{c}{Predicted landing site} & Error in \\
~ & ~ & ~ & Lat. & Lon. & Altitude & Lat. & Lon. & flight & range $d$ & ~ & Lat. & Lon. & prediction \\
\multicolumn{2}{|c}{[GMT]} & ~ & [degrees] & [degrees] & [m] & [degrees] & [degrees] & [minutes] & [km] & [hours] & [degrees] & [degrees] & $\Delta r$ [km] \\
\hline
~2018-03-04* & 10:11 & Switzerland & 46.72 & 6.56 & 26091 & 46.98 & 7.12 & 65.0 & 51.7 & 4.5 & 46.91 & 7.0 & 12.08 \\
2018-04-06 & 08:54 & Switzerland & 46.62 & 7.0 & 28988 & 46.63 & 7.3 & 54.0 & 22.8 & 9.2 & 46.62 & 7.3 & 1.38 \\
2018-04-06 & 18:36 & Switzerland & 46.59 & 6.89 & 23178 & 46.59 & 7.09 & 54.0 & 15.1 & 6.9 & 46.59 & 7.07 & 2.11 \\
2018-05-06 & 07:47 & Switzerland & 46.63 & 7.05 & 25895 & 46.57 & 6.79 & 55.0 & 20.8 & 8.1 & 46.6 & 6.8 & 3.53 \\
2018-05-11 & 10:18 & Switzerland & 46.61 & 6.88 & 28932 & 46.65 & 6.98 & 52.2 & 8.4 & 4.6 & 46.64 & 6.96 & 1.69 \\
2018-05-19 & 08:35 & Switzerland & 46.52 & 6.74 & 28426 & 46.47 & 6.89 & 54.0 & 12.1 & 8.9 & 46.48 & 6.88 & 1.18 \\
2018-06-16 & 11:11 & Switzerland & 46.68 & 6.77 & 21803 & 46.67 & 6.95 & 67.1 & 13.4 & 5.5 & 46.67 & 6.94 & 0.98 \\
2018-06-30 & 19:23 & Switzerland & 46.69 & 6.67 & 25882 & 46.6 & 6.88 & 48.3 & 19.2 & 7.7 & 46.61 & 6.86 & 2.23 \\
~2018-08-03* & 19:06 & Greenland & 67.75 & -48.52 & 28557 & 67.97 & -48.31 & 48.0 & 26.1 & 7.4 & 67.95 & -48.28 & 2.41 \\
2018-08-06 & 07:24 & Greenland & 69.31 & -50.81 & 22134 & 69.34 & -50.56 & 51.0 & 10.3 & 7.7 & 69.33 & -50.57 & 1.04 \\
2018-08-11 & 11:00 & Greenland & 69.74 & -51.2 & 8765 & 69.87 & -51.16 & 42.2 & 15.1 & 5.2 & 69.86 & -51.16 & 1.14 \\
2018-08-13 & 17:00 & Greenland & 69.43 & -51.52 & 21457 & 69.56 & -51.63 & 42.3 & 15.1 & 5.3 & 69.57 & -51.64 & 1.54 \\
2018-10-28 & 11:30 & Morocco & 30.78 & -5.61 & 25892 & 30.78 & -4.85 & 45.0 & 72.8 & 5.8 & 30.79 & -4.83 & 1.92 \\
2018-11-17 & 10:24 & Switzerland & 46.36 & 7.54 & 24621 & 46.09 & 6.98 & 44.0 & 51.8 & 4.7 & 46.12 & 6.99 & 2.65 \\
2018-12-15 & 10:11 & Switzerland & 45.85 & 7.9 & 26435 & 45.39 & 8.36 & 51.0 & 62.6 & 4.5 & 45.41 & 8.37 & 2.59 \\
2019-01-01 & 13:41 & Switzerland & 45.78 & 8.18 & 25863 & 45.39 & 8.43 & 52.0 & 47.1 & 8.0 & 45.4 & 8.43 & 0.85 \\
2019-02-02 & 13:30 & Switzerland & 46.58 & 6.67 & 20706 & 46.79 & 6.79 & 45.0 & 24.6 & 7.8 & 46.74 & 6.76 & 5.74 \\
2019-02-09 & 15:30 & Switzerland & 45.94 & 8.16 & 14030 & 45.85 & 8.69 & 33.0 & 42.4 & 3.8 & 45.85 & 8.71 & 1.77 \\
2019-03-02 & 12:24 & Switzerland & 46.01 & 6.83 & 25950 & 45.98 & 7.0 & 39.0 & 13.7 & 6.6 & 45.96 & 7.03 & 3.46 \\
2019-03-11 & 17:53 & Switzerland & 45.11 & 7.31 & 27358 & 44.42 & 7.79 & 48.0 & 85.8 & 6.2 & 44.37 & 7.83 & 6.28 \\
2019-04-19 & 18:00 & Switzerland & 31.27 & -6.86 & 23546 & 31.15 & -6.27 & 44.5 & 56.8 & 6.3 & 31.15 & -6.26 & 1.75 \\
2019-04-21 & 18:18 & Switzerland & 31.25 & -6.96 & 24493 & 30.9 & -6.61 & 42.0 & 50.3 & 6.6 & 30.93 & -6.6 & 2.88 \\
2019-04-22 & 13:30 & Switzerland & 30.78 & -7.42 & 25594 & 30.57 & -7.14 & 48.0 & 35.0 & 7.8 & 30.61 & -7.15 & 4.28 \\
2019-04-24 & 14:30 & Switzerland & 31.26 & -6.91 & 23507 & 31.17 & -6.57 & 39.0 & 33.5 & 8.8 & 31.18 & -6.55 & 2.61 \\
2019-04-25 & 13:41 & Switzerland & 31.25 & -9.35 & 24881 & 31.3 & -9.11 & 45.5 & 22.8 & 8.0 & 31.31 & -9.11 & 1.47 \\
~2019-05-31* & 12:11 & Morocco & 46.06 & 6.02 & 23838 & 45.95 & 5.99 & 30.0 & 12.4 & 6.5 & 45.88 & 5.97 & 7.81 \\
2019-07-26 & 14:30 & Switzerland & 32.03 & -6.5 & 28321 & 31.94 & -6.45 & 42.2 & 10.6 & 8.8 & 31.93 & -6.48 & 2.2 \\
2019-07-28 & 19:53 & Switzerland & 31.36 & -9.31 & 27829 & 31.43 & -9.23 & 51.0 & 11.3 & 8.2 & 31.42 & -9.22 & 1.78 \\
2019-07-30* & 08:11 & Switzerland & 29.52 & -9.94 & 23879 & 29.56 & -9.82 & 30.0 & 12.2 & 8.5 & 29.59 & -9.8 & 4.06 \\
2019-08-01 & 07:00 & Switzerland & 31.27 & -4.13 & 25439 & 31.47 & -3.97 & 48.0 & 26.0 & 7.3 & 31.45 & -4.0 & 3.41 \\
\hline
2019-09-18 & 02:58 & Canada & 47.76 & -80.72 & 28400 & 47.49 & -80.56 & 38.6 & 32.4 & 9.2 & 47.5 & -80.56 & 0.54 \\
2019-09-18 & 17:52 & Canada & 47.49 & -82.4 & 29848 & 47.42 & -82.17 & 35.4 & 18.6 & 6.1 & 47.42 & -82.19 & 1.12 \\
\hline
\end{tabular}
\label{tab:funkflights}
\end{sidewaystable}

\subsection{Method: Dynamical Modelling}
\subsubsection{Initial Conditions}\label{sec:initialconditions}

The user inputs the starting location, $\mathbf{r}^{\rm release}$=(longitude, latitude) and altitude $z$, as well as the date and time of release (this defaults to now). If desired, a `drift time' can be specified, during which the \drs travels horizontally with the \hab platform before release. 
The code automatically determines and downloads the most appropriate GFS weather data for these inputs.

Upon release, we assume that the \drs instantly reaches terminal velocity. Balancing gravitational acceleration $g$ acting downwards and drag force acting upwards, this is
\begin{equation}\label{eq:vd}
v_{z}^{\rm predicted} = - \lambda \left(\frac{m}{A\,C_{d}}\right)^{\frac{1}{2}}\left(\frac{2g}{\rho}\right)^{\frac{1}{2}},
\end{equation}
where $m$ is the mass of the payload, $\rho$ is the density of air, $A$ is the area of the parachute, and $C_{d}$ is its coefficient of drag. We initially adopt the manufacturer's design specifications for $A$ and $C_{d}$ (see section \ref{sec:parachute}), but calibrate these via free parameter $\lambda$ (see section \ref{sec:parachute}). Both $g$ and $\rho$ depend on altitude; we calculate $g(z)$ assuming the Earth is a perfect sphere with a radially symmetric distribution of mass and interpolate $\rho$ from the GFS weather model.

\subsubsection{Iterated Descent Trajectory}\label{sec:motion}

We split the descent into altitude steps of height $\Delta z$ {(we set a requirement on this in section \ref{sec:weather_data})}. 
For each altitude step, we calculate the time $\Delta t$ to descend from top to bottom, assuming that the parachute moves vertically with the terminal velocity evaluated at the midpoint of the altitude step, directly below its starting position. The main strength of this 'leapfrog' method of updating the velocity is that it better conserves the energy of the dynamical system and therefore does not allow the system to drift substantially over time. By using this method, we better approximate the true velocity versus altitude curve than if instead we used the velocity at the beginning of the altitude step.

We neglect updraughts and downdraughts in the GFS model, finding these negligible to the terminal velocity and having no measurable effect on the accuracy of our predicted landing sites.

During each altitude step, we assume that the parachute and payload travel horizontally with North-South (`$u$') and East-West (`$v$') wind speeds, again evaluated directly below the starting position, at the midpoint of the altitude step.
We update the latitude and longitude of the \drs using the haversine formula, then iterate to the next altitude step.

\subsubsection{Termination Criterion}

The code iterates the position of the \drs until it reaches sea level (altitude $z=0$). This is generally below ground. We do not test for this during descent, because calls to evaluate ground level are relatively slow, and fast horizontal speeds near the ground necessitate a new call at each step.\footnote{Checking that the \drs is above ground at each time step adds 1\,s to runtime if $\Delta z=100$\,m, or 20\,s for $\Delta z=1$\,m.} We instead work backwards from $z=0$, checking whether each point in the predicted trajectory was above or below ground. Once we find a pair of coordinates straddling ground level, we interpolate linearly between them to predict the latitude and longitude of the landing site, $\mathbf{r}$. 

\begin{figure}[t]
\centering
\includegraphics[width=0.8\textwidth]{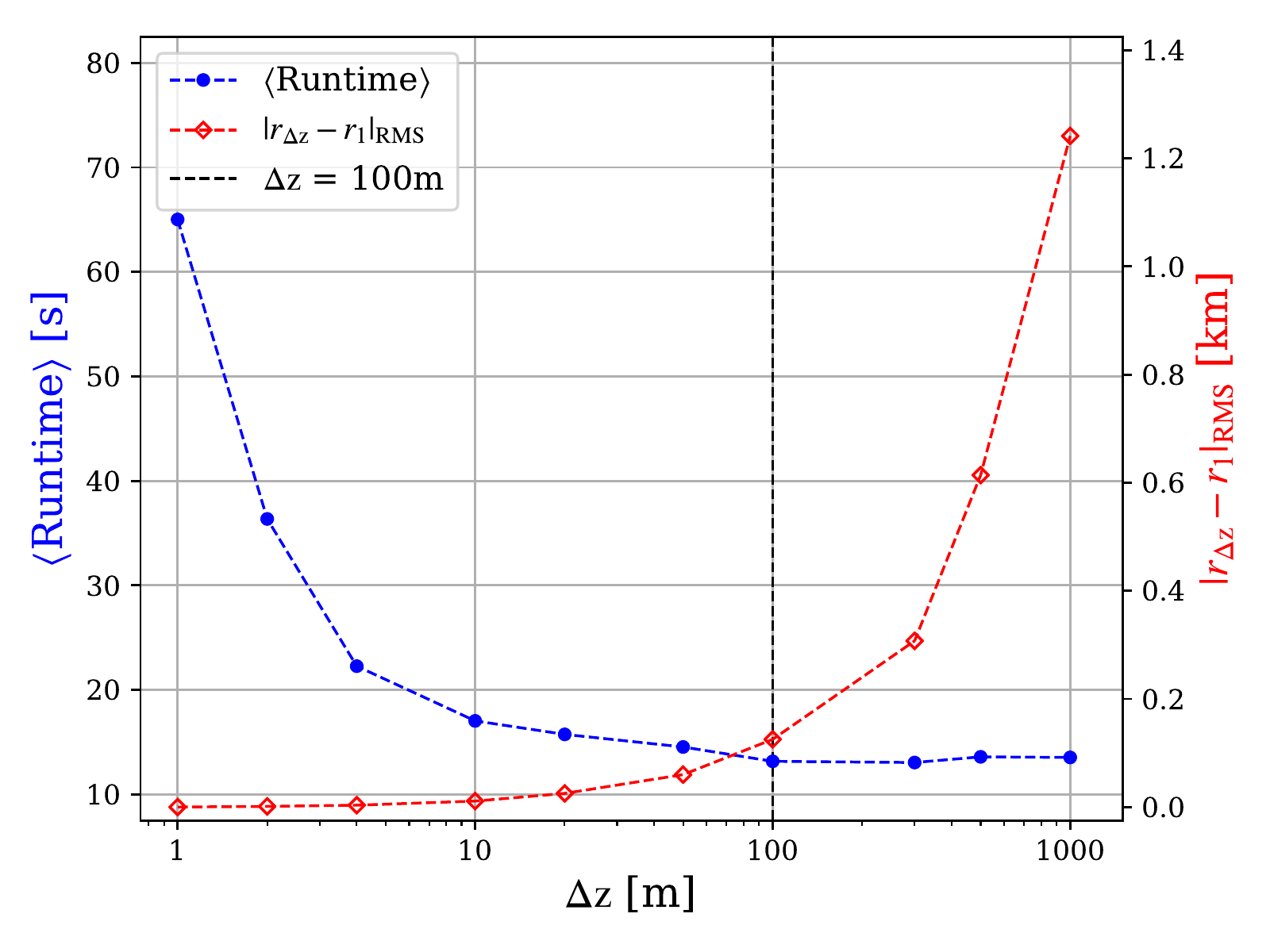}
\caption{
Code convergence test, and tradeoff between precision versus speed.
{\emph Red}: the root mean square horizontal error in predicted landing site as a function of altitude step size $\Delta z$, compared to the most accurate prediction using $\Delta z=1$\,m. 
{\emph Blue}: mean wallclock runtime per trajectory calculation, on a 1.7GHz laptop.
In both cases, trajectories are calculated from, and averaged over all 30 initial conditions in table~\ref{tab:funkflights}.
The vertical dashed black line indicates our choice of nominal altitude step $\Delta z=100$\,m that is used for all further analysis in this paper.}
\label{fig:altstep}
\end{figure}

\subsubsection{Convergence Test}

The choice of altitude step size $\Delta z$ represents a tradeoff between precision and run-time.
Run-time is important for real-time predictions of the landing site, to optimise the moment of release from a fast-moving \hab (requirement  R7).
To predict the landing site $\mathbf{r}_{1}$ with the greatest possible precision (but slowly), we use altitude step size $\Delta z=1$\,m to calculate trajectories from all the initial conditions in table~\ref{tab:funkflights}, as a representative sample of possible release locations.
We then recompute the trajectories with different step sizes, and record predicted landing sites $\mathbf{r}_{\Delta z}$.
The mean error $\langle\mathbf{r}_{\Delta z}-\mathbf{r}_\mathrm{1m}\rangle$, and the wall-clock runtime on a laptop with a 1.7\,GHz CPU are shown in figure~\ref{fig:altstep}. Note that during calculation of the trajectories, we did not check for ground elevation.

Predictions for the landing site converge successfully if the altitude step size fully samples the (maximum 200\,m) vertical resolution of the GFS models. 
A practical compromise is $\Delta z=100$\,m.
In a runtime of 13\,seconds, this achieves a mean landing site precision of 125\,m: an error that is subdominant to other sources of uncertainty. 
All further analysis will be performed with this step size.

\begin{figure}[t]
    \centering
    \includegraphics[width=0.8\textwidth]{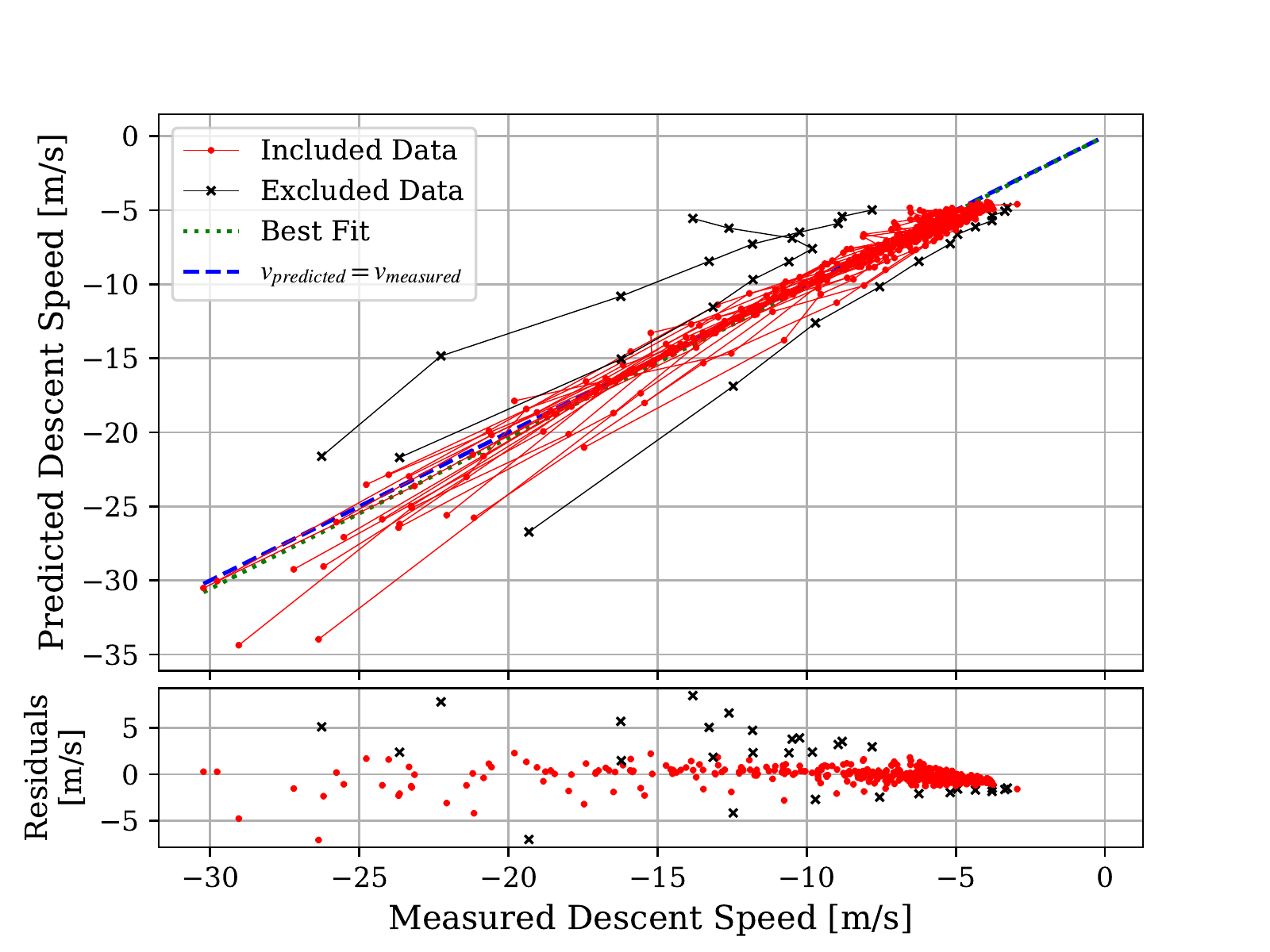}
    \caption{Calibration of our parachute descent model, by comparing (only) the vertical speed predicted for and recorded during the 30 test flights.
    \emph{Top panel}: the descent speed for each flight {(red and black lines for included and excluded flights respectively)};  trajectories start on the left and end on the right, with data points recorded every $\sim$5\,minutes.
    If our trajectory calculation were perfect, the predicted and actual descent speeds would be equal (blue dashed line). The best-fit linear perturbation from this is consistent with the speeds having been overestimated by ($3.7\pm0.4$)\% (green dotted, which is constrained to pass through the origin). 
    \emph{Bottom panel}: residuals of the best fit to the data in the top panel.}
    \label{fig:vdes}
\end{figure}

\subsection{Trajectory Calibration and Validation}
\subsubsection{Vertical Descent Speeds}\label{sec:vspeeds}

We compare the vertical component of the predicted descent speeds to the altitude difference between successive \gnss measurements, for 29 of the 30 test flights (figure~\ref{fig:vdes}).\footnote{The \gnss failed to record during most of the 2018-08-03 flight in Greenland (most likely due to cold), so we exclude this flight from figure~\ref{fig:vdes} and all subsequent analysis.} 
The predicted and measured speeds would be equal, if the design specification of the parachute's drag coefficient and area were correct, and the payload masses were recorded correctly.
To refine our knowledge of these parameters, we fit the free parameter $\lambda$ from equation~\eqref{eq:vd} across all flights, as
\begin{equation}\label{eq:vd_fit}
v_z^{\rm predicted} = \lambda v_z^{\rm measured}. 
\end{equation}
The best-fit value is $\lambda_\mathrm{bf}=1.019\pm0.006$. 
There is a marginal evidence that the predicted speeds are approximately correct at high speed (high altitude), but 10--20\% too low at low speed (low altitude). 
This might be due to additional drag in the higher density air -- but without further evidence to support and quantify this hypothesis, we shall consider it useful margin in safety requirement \ref{req:vertical_speed}, and empirically incorporate it into our uncertainty in the predicted landing sites.

In our test data, the payload mass and parachute diameter were not precisely recorded.
To test whether these varied between flights, we refit $\lambda$ for each individual flight. 
Three flights in particular (2018-03-04 in Switzerland, 2019-05-31 and 2019-07-30 in Morocco) have large ($>4$\,km) errors in their predicting landing sites (see table \ref{tab:funkflights}) and also have the most anomalous values of $\lambda_\mathrm{bf}$. 
They are so different from $\lambda_\mathrm{bf}=1$ that either $m<1$\,kg (unlikely for practical reasons), $m>2$\,kg (impossible for legal reasons), or (most likely) a different parachute was used. 
We exclude these three flights from further quantitative analysis. All other 26 test flights have descent rates consistent with a mean value of $\langle 1/\lambda_\mathrm{bf}\rangle=0.967\pm0.005$. Individual values of $\lambda_\mathrm{bf}$ vary by $<20\%$; if we use these values to recompute the trajectory, the mean error in landing site (compared to the truth) changes negligibly from 2.40\,km to 2.37\,km. We thus conclude that both the parachutes and payload masses were likely constant for these flights. Nonetheless, because $\lambda_\mathrm{bf}$ is always consistent with 1, yet the true payload mass remains uncertain, we henceforth adopt $\lambda=1$ for all further calculations. 
If the payload masses did vary between flights, this approach will lead to a slight increase in our estimate of uncertainty. However, it should avoid biasing the calculation of future trajectories with different payload masses.

\subsubsection{Horizontal Position}\label{sec:errs}

\begin{figure}[t]
    \centering
    \includegraphics[width=\textwidth]{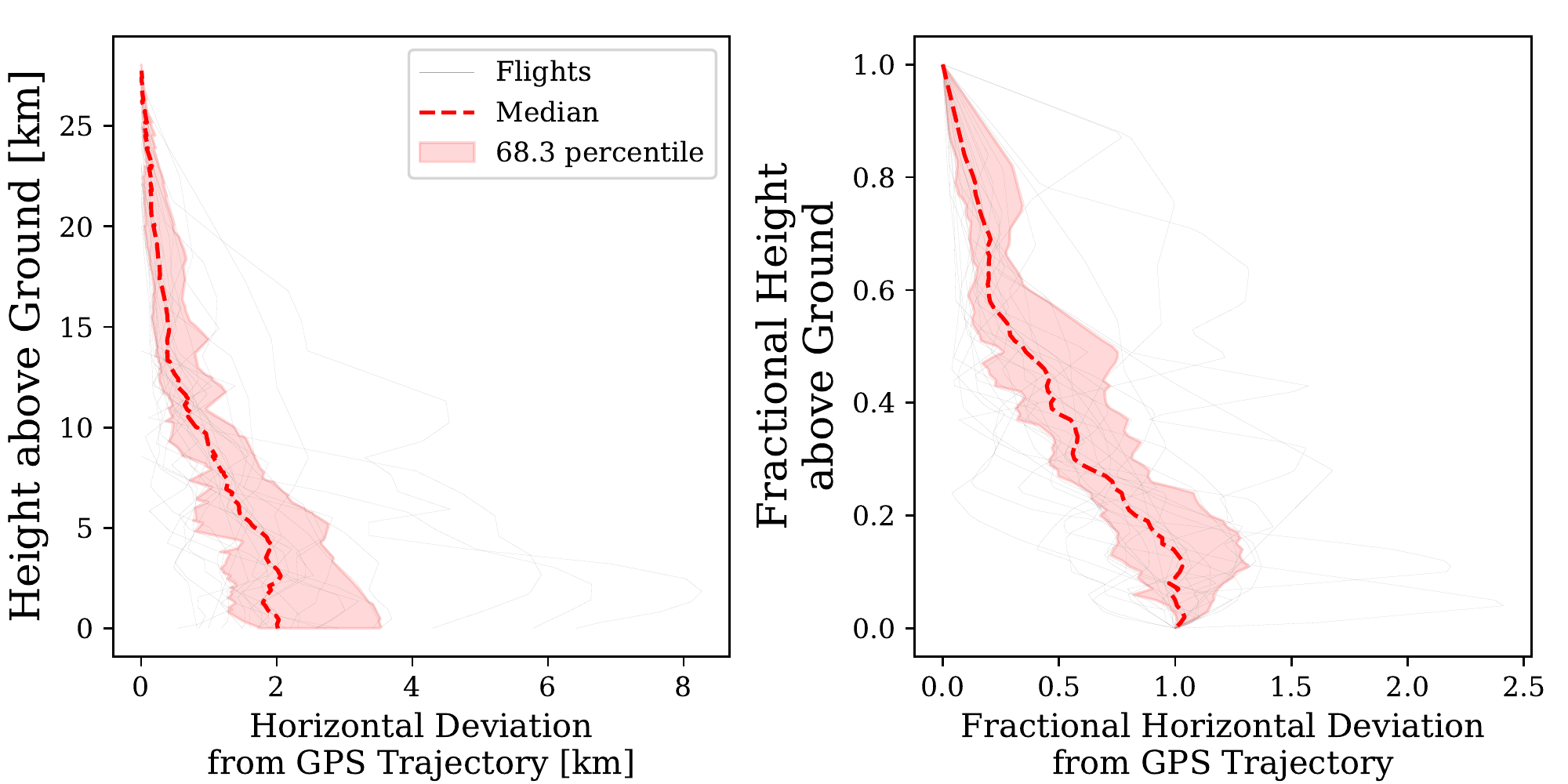}
    \caption{Accuracy of trajectories predicted for the descent of 26 parachutes, compared to the true trajectories recorded by \gnss. Trajectories begin at the top, and end at the bottom.
    \emph{Left panel}: absolute horizontal deviation of each true trajectory from the prediction, at heights above ground level whenever the \gnss location was recorded, every $\sim$5\,minutes. Each descent begins from a slightly different altitude. The red line indicates the median of the 26 flights, and the red area indicates the 68.3\% region.
    \emph{Right panel}: as before, but with the vertical and horizontal distance covered by each trajectory normalised to start or end at the same fractional altitude or horizontal deviation.}
    \label{fig:alt_vs_z}
\end{figure}

The most important aspect of a predicted trajectory is its horizontal accuracy, which culminates in the distance of its predicted landing site from the true landing site, $\Delta \mathbf{r}=(\mathbf{r}_{\rm predicted}-\mathbf{r}_{\rm true})$. 
We find that our predicted trajectories are most accurate at high altitude, which is traversed quickly, and near the ground, where the weather forecast is higher resolution and perhaps more accurate (figure~\ref{fig:alt_vs_z}). 

Most of the deviation from the predicted trajectory builds while the parachute descends through the jet stream, where horizontal speeds are also greatest. 
Thus, the accuracy of our predictions is probably more limited by the accuracy of weather forecasts than the accuracy of our time-stepping algorithm.

We model uncertainty in the predicted landing site as 
\begin{equation}\label{eq:sigmas}
\sigma_{\parallel}^{2} = (q\sigma_{\perp})^{2} \equiv \sigma_{0}^{2} + hd_{\rm predicted}^{2} + k\langle t_{\rm future}\rangle^{2},
\end{equation}
where $d_{\rm predicted}$ is the horizontal distance between the release point and predicted landing site, $\langle t_{\rm future}\rangle$ the average $t_{\rm future}$ of the forecasts used at each altitude step in a predicted trajectory --- and $\sigma_{\parallel}$, $\sigma_{\perp}$, $q$, $\sigma_{0}$, $h$ and $k$ are free parameters.
In particular, $\sigma_{\parallel}$ ($\sigma_{\perp}$) is our model uncertainty in (perpendicular to) the mean direction of predicted travel, and $q$ is the axis ratio between them. 

We fit the free parameters using Python code {\sc emcee} \citep{j} to maximise log-likelihood
\begin{equation}\label{eq:likelihood}
\ln{\mathcal{L}} \equiv -\frac{1}{2}\sum_{i=1}^{26}\left[(\Delta r_{\parallel,i} - \sigma_{\parallel,i})^{2}+(\Delta r_{\perp,i} - \sigma_{\perp,i})^{2}\right],
\end{equation}
where $\Delta r_{\parallel,i}$ ($\Delta r_{\perp,i}$) is the component of $\Delta \mathbf{r}$ in (perpendicular to) the direction of $d_{\rm predicted}$, for each descent in table~\ref{tab:funkflights}.
We compute two sets of predicted trajectories.
The first set is relevant to assess the safety and optimum timing of a live release, and uses only those weather forecasts that would be available at release (or earlier, to constrain $k$).
The second set is the most accurate that could be made to aid recovery, if communications were lost with {\sc DRS} capsules immediately after release. These interpolate between weather forecasts available before and after launch, and also use $\Delta z=1$\,m, for a slower but slightly more accurate calculation.

\begin{table}[t]
\centering
\caption{Best-fit parameters for model \eqref{eq:sigmas} of the uncertainty in predicted landing sites, after predicting all the descents in table~\ref{tab:funkflights}. The two sets of parameters represent predictions made using only those weather forecasts available before release, or also those spanning the time of release and available shortly after.}
\smallskip
\begin{tabular}{|ccccc|} 
\hline
\multirow{2}{*}{Weather forecast models} & $\sigma_{0}$ & $h$ & $k$ & $q$ \\
 & [km] & [$10^{-4}$] & [$\rm 10^{-3}\,km^{2}/hour^{2}$] & ~ \\
\hline
Available at launch & $1.77\pm0.14$ & $3.1\pm1.5$ & $3.6\pm0.9$ & $1.14\pm0.06$ \\
Available with hindsight & $1.63\pm0.13$ & $6.4\pm1.6$ & $3.3\pm1.1$ & $1.20\pm0.07$ \\
\hline
\end{tabular}
\label{tab:table2}
\end{table}

In both cases, the uncertainty is slightly greater in the direction of travel ($q>1$); we convert the best-fit parameters into error ellipses on the predicted landing sites.

\section{End-to-End System Test}\label{sec:test}

We shall now describe an end-to-end test of the \drs hardware and software performed during the 2019 science commissioning flight of the \superbit telescope. In general, \drs capsules could be released at any time during a \hab mission, with only a few minute's notice. For convenient retrieval, we planned to release one \drs shortly after reaching ceiling (so that it would land near the launch base) and the second shortly before termination (so that it would land near the main gondola). To save cost, the \drs capsules were configured for this test with only 1\,TB of storage ($1\times512$\,GB plus $4\times128$\,GB) instead of the maximum 5\,TB.

\subsection{Launch and Release}\label{sec:launch}

\begin{figure}[h]
\centering
\includegraphics[width=0.71\textwidth]{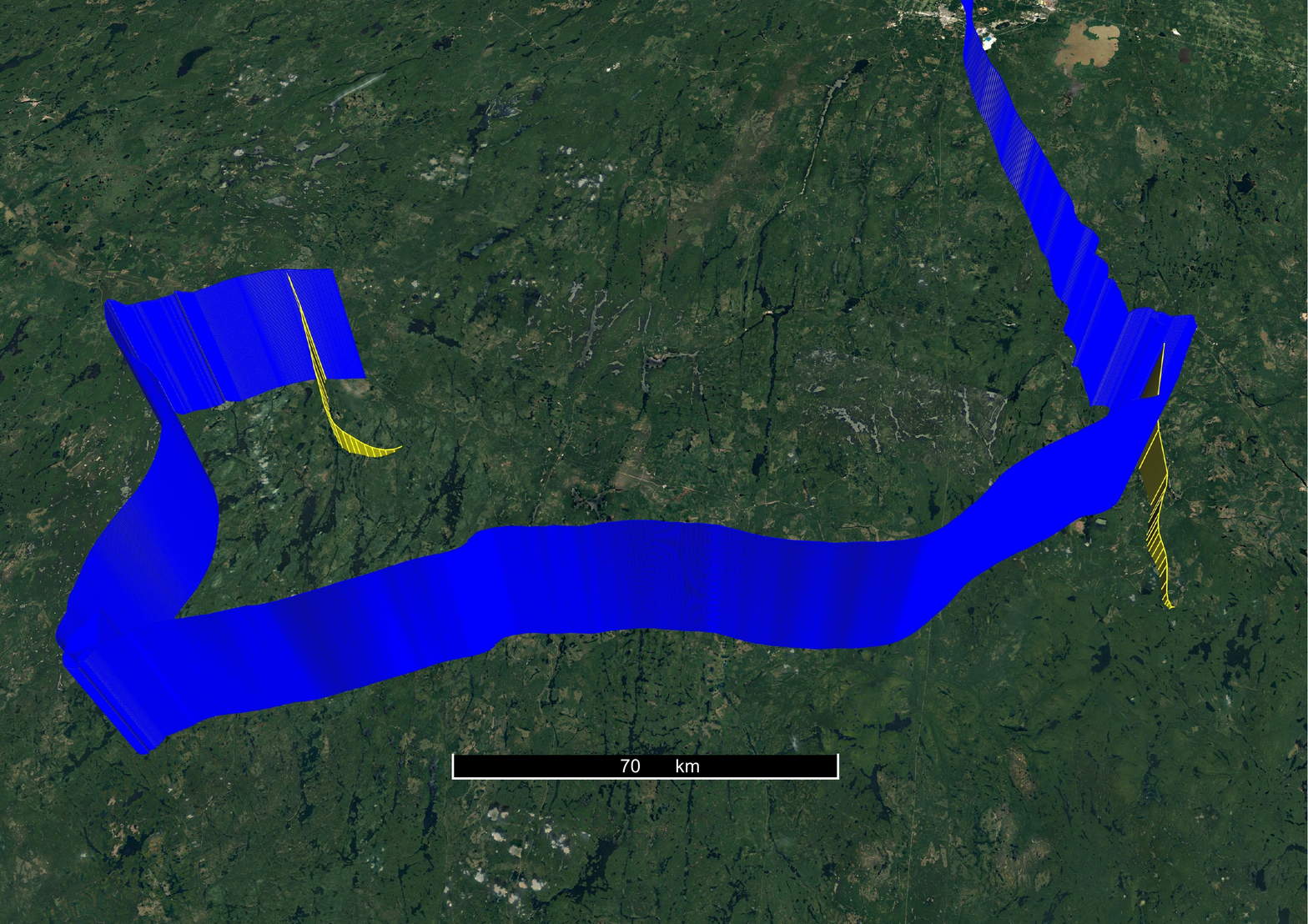}
\caption{The flight path of the two \drs capsules, while they were attached to \superbit (blue) and while descending independently by parachute (yellow). The trajectory starts near the top right corner of the figure, and continues clockwise. It does not include \superbit's descent because the main gondola powers down before termination.}
\label{fig:superbit_traj}
\end{figure}

The \superbit telescope was launched from the \cnes Stratospheric Launch Base in Timmins, Ontario on 2019-09-17 at 20:34 GMT-4, carrying two \drs capsules (figure~\ref{fig:superbit+capsules}).
During ascent, we obtained science calibration data from the telescope, and copied it to the \drs capsules.
Shortly after ascent through $\sim$28\,km altitude, we used our trajectory prediction software to target an area of forest without lakes or population, yet still near enough to the launch facility for convenient retrieval.
We waited until the \drs would land near remote but usable roads identified in satellite imagery, then released the first \drs capsule with predicted $1\sigma$ uncertainties on the landing site of $2.0$\,km and $1.7$\,km in the directions parallel and perpendicular to the direction of travel respectively. 

The \superbit mission continued, performing telescope calibration and alignment -- followed by 3.5\,hours acquiring science data that was copied to the second \drs.
We planned to release the second \drs shortly before mission termination, so that it would land near the \superbit gondola, convenient for retrieval.
In the event, the mission was terminated early because \superbit's balloon had a leak.
We still released the \drs shortly before termination but, because of time constraints, did not have opportunity to run our prediction software in advance. This was acceptable from a safety perspective because the main gondola was predicted (by proprietary \cnes software) to land well away from population, and had a similar value of $m/AC_d$ as the \drs.
We released the \drs, and afterwards ran our prediction software for the moment of release, using weather forecasts that would have been available in advance.
{Predicted $1\sigma$ uncertainties on the landing site were $1.9$\,km and $1.6$\,km in the directions parallel and perpendicular to the direction of travel respectively.}

Figure~\ref{fig:superbit_traj} shows the full trajectory of \superbit, recorded by its own \gnss receiver, and the trajectories of both \drs capsules. Coordinates of the \drs release points are included in table~\ref{tab:funkflights}.

\begin{figure}[t]
\centering
\includegraphics[width=0.95\textwidth]{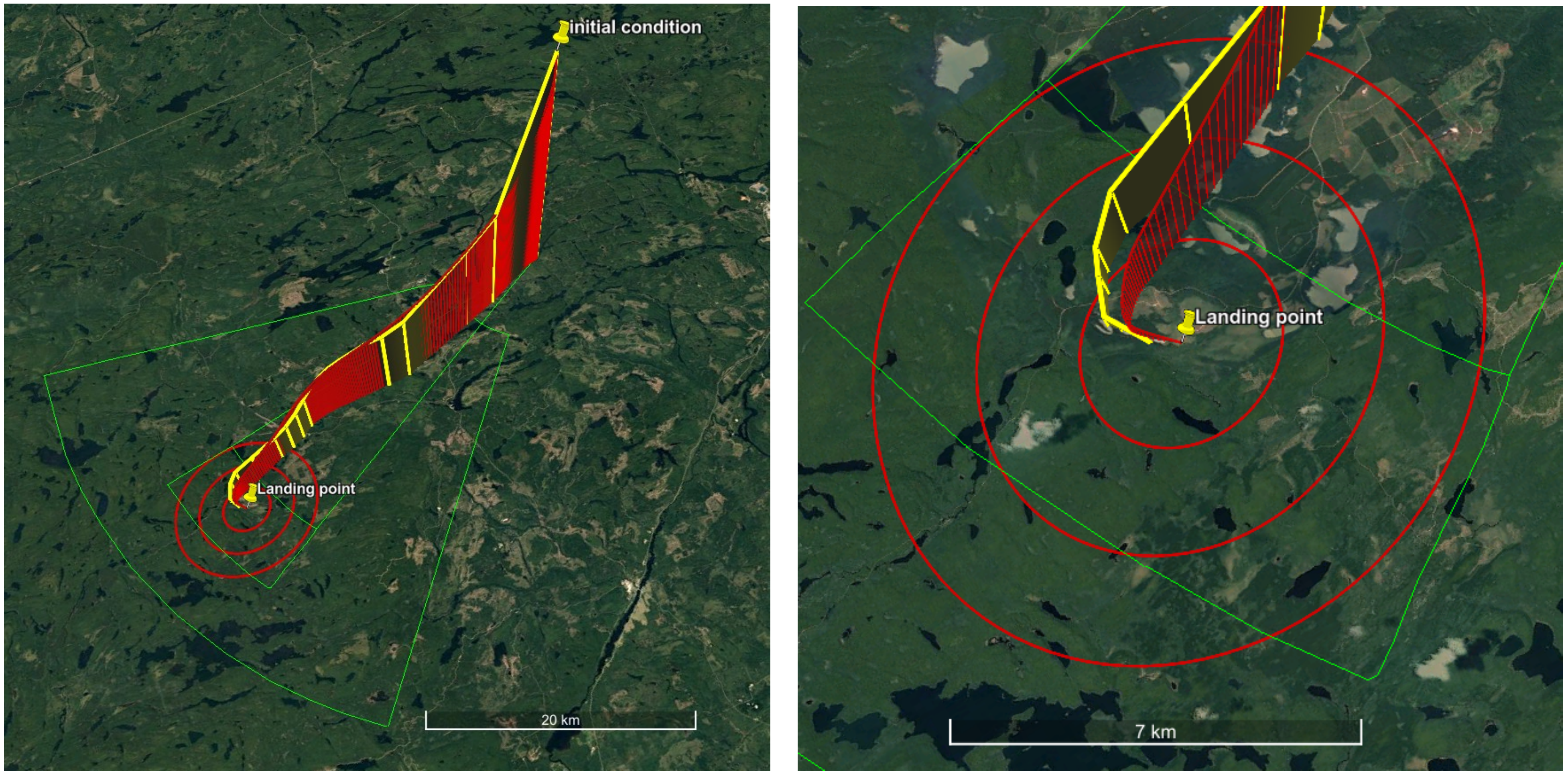}
\caption{The predicted trajectory of the first \drs capsule, using GFS weather forecast data available at launch (red), and its actual trajectory recorded by \gnss (yellow). The yellow pin labeled `initial condition' on the top right marks its release location. The yellow pin labeled `Landing point' marks its predicted landing location, surrounded by red ellipses indicating 1, 2, and 3$\sigma$ uncertainty. Narrow and wide green cones show the 1 and 3$\sigma$ predictions from \cnes software. The right panel is a zoom of the left.}
\label{fig:cap1}
\end{figure}

\begin{figure}[t]
\centering
\includegraphics[width=0.95\textwidth]{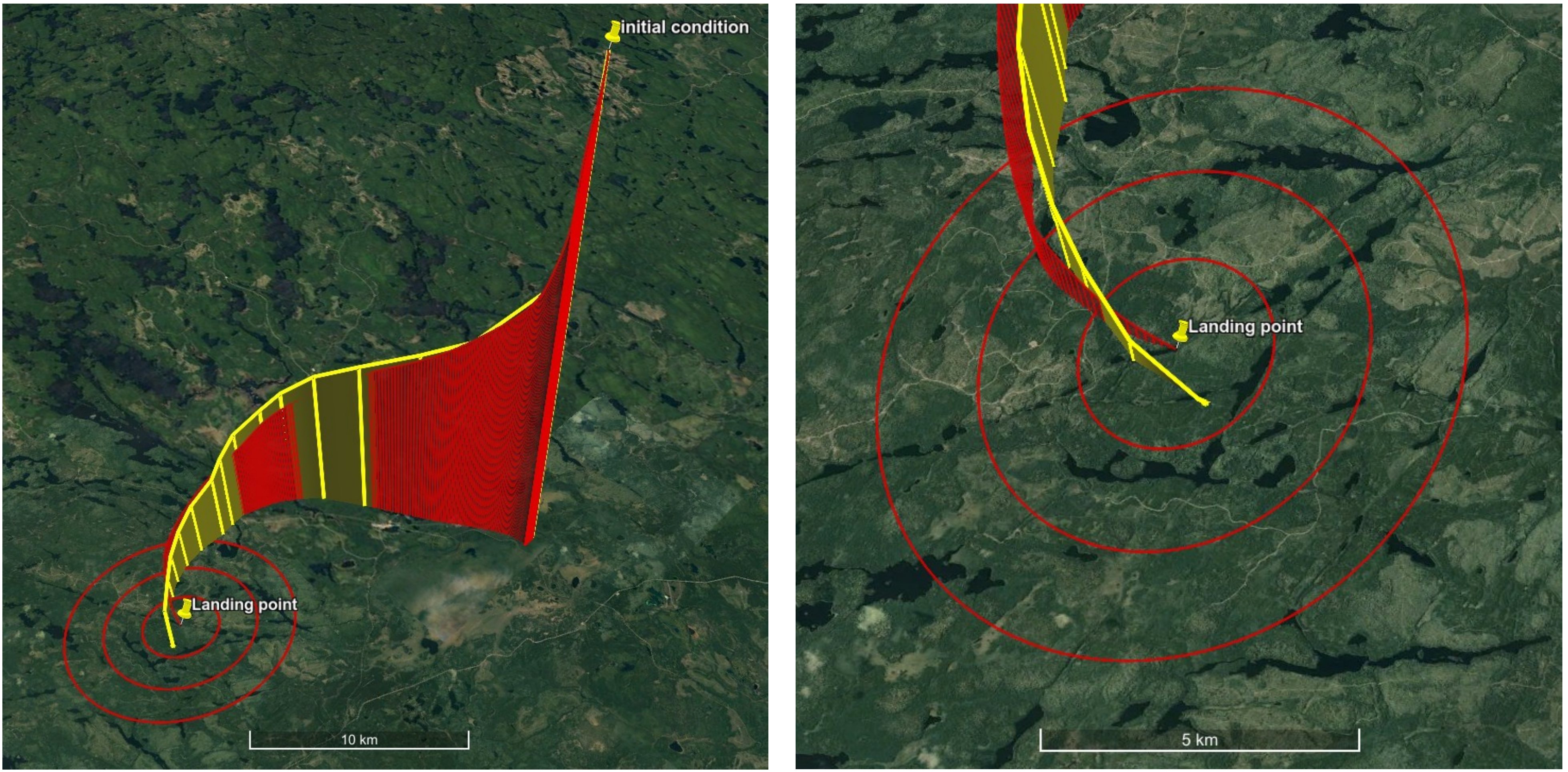}
\caption{As figure \ref{fig:cap1}, but showing the predicted (red) and \gnss (yellow) descent trajectory of the second \drs capsule.
The prediction from the \cnes software was used before dropping the capsule, but is no longer available for inclusion in this figure.
}
\label{fig:cap2}
\end{figure}

\subsection{Descent and Landing}\label{sec:res}

Both \drs capsules began logging \gnss coordinates before release, and continued transmitting them via Iridium, every $\sim$2\,minutes (17 and 20 times) during descents lasting 35 and 39\,minutes. We had increased the frequency of these transmission for better localisation in case of lost contact, because of high winds at ground level that week. Indeed, western Canada is covered by dense forest \citep{k}, so \gnss lock from the forest floor was not guaranteed.

Both capsules maintained Iridium link after landing, and continued reporting \gnss coordinates with standard deviation in latitude and longitude of 7\,m from the first \drs, and 10\,m from the second. We waited to receive a few dozen \gnss readings, to average away this noise, then commanded the capsules via Iridium MT message to conserve battery life and report back only every 2\,hours. Both capsules had landed safely, on dry land.

The predicted trajectories were more accurate than expected (figures~\ref{fig:cap1} and \ref{fig:cap2}). Predicted landing sites were within 300\,m and 600\,m of the true locations, which would have been adequate for successful recovery even without \gnss measurements. We obtained live predictions using an older version of the software than that available on github.\footnote{For example, the `leapfrog' method of updating the position and velocity discussed in section \ref{sec:motion} was not implemented in the older version of the code.} The current version is more accurate in general but -- for these particular initial conditions -- predicts landing sites within 600\,m and 1100\,m of the true locations, consistent with the expected uncertainty. Our live runs were noisier, and their particularly high accuracy was good luck. 

\begin{figure}[t]
\centering
\includegraphics[width=0.95\textwidth]{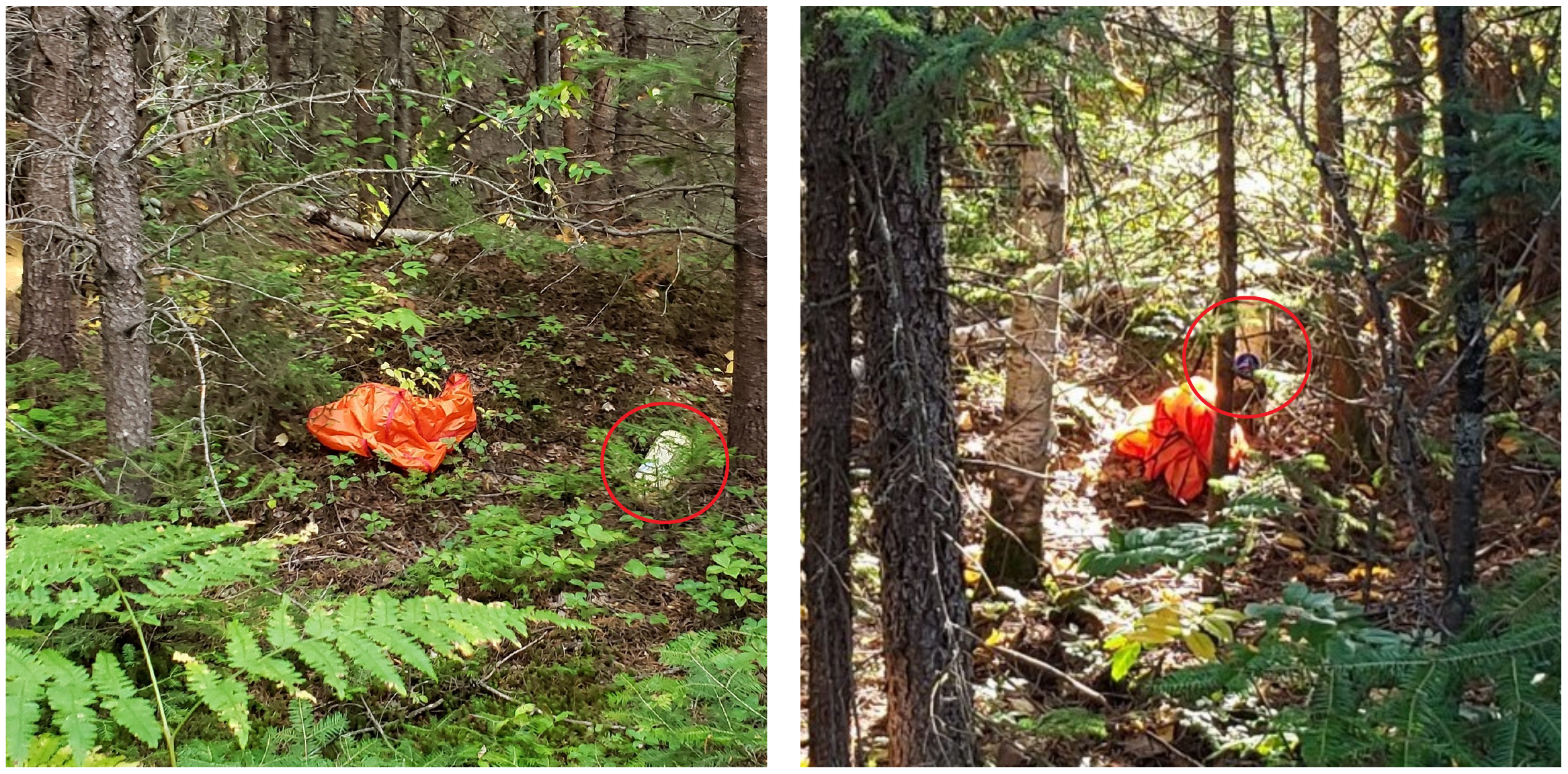}
\caption{Photos of the two capsules on the ground taken by the \cnes recovery team S\'ebastian Lafrance and Francis Martin. The capsules are indicated by red circles. The parachutes can be clearly seen in bright orange.}
\label{fig:recovery}
\end{figure}

\subsection{Recovery}

To aid recovery, the capsules are equipped with a sounder, and the parachutes are bright orange.
A recovery crew went to the \gnss coordinates of both landing sites, and found both \drs capsules within a few minutes each.
They had both fallen to the forest floor (figure~\ref{fig:recovery}), so no further action was necessary.

Upon return to the launch facility, the cases were opened to remove batteries and deactivate the sounders (they could have been deactivated remotely, but were in the back of an effectively soundproof truck).
A few pine needles had entered the upper chamber of one \drs, but the inner chamber of both \drs capsules was clean.
The raspberry Pis were plugged into external power, and the data successfully retrieved.

\section{Conclusions}\label{sec:conclusion}

Retrieving assets from a High Altitude Balloon (\hab) platform can mitigate the risk of total loss if the platform is damaged or lost upon landing. 
Mid-flight retrieval can also increase a mission's efficiency, if its initial performance is assessed, and subsequent operation improved. 
One solution to retrieve physical samples, or digital data acquired at too high a rate for transmission to the ground, is to jettison a small capsule that descends via parachute. 

We have developed, and successfully tested the \superbit Data Recovery System (\drs) to `download' up to 5\,TB of data via parachute.
We released two \drs capsules from $\sim30$\,km altitude during a commissioning flight of the \superbit telescope in September 2019. \superbit is an astronomical telescope that operates in the stratosphere for up to 100\,days at a time.  
Both capsules landed safely, a few hundred metres from their predicted landing sites, and were easily recovered.\vspace{2mm}

Hardware worked as envisaged. 
Several times during flight, the main gondola logged in to the \drs capsules via 2.4\,GHz Wi-Fi$^{\rm TM}$, and copied data onto them. 
At two different times, we issued a two-stage `release' command to one \drs, via {\tt ssh}.
The capsules dropped 30\,seconds later, and their parachutes opened. During and after descent, they measured their location via \gnss and transmitted it back to the ground station via Iridium message.

Software to predict the descent trajectory also worked well. After travelling a horizontal distance of 31 and 19\,km from their release points, the \drs capsules landed within 300\,m and 600\,m of their expected landing sites. Calibrated on 30 parachute descents from the stratosphere, our software can predict landing sites all over the world with 1$\sigma$ uncertainty of $\sim$1.5\,km. This uncertainty accumulates most rapidly while the capsules descend through the jet stream. Our software thus appears limited mainly by the accuracy of (GFS) weather models at this altitude. Nonetheless, it satisfies safety requirements to permit immediate release --- and it can also be used to predict the best time to release a capsule so that it can be conveniently recovered. This takes the form of a landing strip on the ground, roughly underneath the future path that the software predicts for the \hab platform. \vspace{2mm}

For future flights, we are considering hardware upgrades including\vspace{-2mm}
\begin{enumerate}
\item[-] Wired ethernet, for faster data transfer, and to avoid any potential for radio frequency (RF) electromagnetic interference. We have flown 2.4\,GHz Wi-Fi$^{\rm TM}$ networks on both \nasa and \csa/\cnes balloons without any problems, but testing for that interference has frequently slowed payload integration, and has even delayed launch on one occasion. Additionally, this would extend the possible applications for the \drs to Cosmic Microwave Background (CMB) experiments, \citep[e.g.\ SPIDER;][]{l}, which are extremely sensitive to RF and would be unable to tolerate an onboard Wi-Fi$^{\rm TM}$ network. 
\item[-] Thermal redesign to enable the Raspberry Pis to pre-process and analyse science data in-flight. If power is abundant, a networked collection of Raspberry Pis represents considerable processing power at the start of a \hab mission, precisely the time when decisions are likely needed to assess data quality and update science targets/goals.
\item[-] An openable and re-sealable chamber to acquire samples of the biome in the upper atmosphere, to be returned for laboratory analysis on the ground.
\end{enumerate}\vspace{-2mm}

\noindent During this test with \superbit, we used the \drs capsules as a means to retrieve digital data. However, we envisage that they could be used to retrieve a variety of assets, including hardware or physical samples. We welcome interest from other \hab teams for whom the system may be useful.

\acknowledgments

The authors offer huge thanks to the mission support teams at \csa and \cnes\ --- especially to Fr\'ed\'eric Thoumieux and Christian Lamarque for help with weather forecasts and parachute dynamics, and to S\'ebastian Lafrance and Francis Martin for hiking through Canadian forest to recover the \drs capsules. We are grateful to Panu Lahtinen for sharing his excellent {\tt pyBalloon} code.

This work was funded by the Royal Society [grants UF150687 and RGF/EA/180026], the UK Science and Technology Facilities Council [grants ST/P000541/1, ST/J004650/1, plus an Impact Acceleration Award to Durham University], and UKRI [grant MR/S017216/1]. Launch and operational support for the 2019 \superbit flight from Timmins, Ontario was provided by the Centre National d'\'Etudes Spatiales (\cnes) and the Canadian Space Agency (\csa). Funding for the development of \superbit is provided by \nasa through {\sc apra} [grant NNX16AF65G]. This work was done in part at JPL, run under a contract for \nasa by Caltech. The Dunlap Institute is funded through an endowment established by the David Dunlap family and the University of Toronto.

\end{document}